\documentclass[aps, prl, amsmath, amssymb]{revtex4}

\begin{document}

\title{Wilson Line Integrals in the Unparticle Action}
\author{A. Lewis Licht}
\affiliation{Dept. of Physics\\U. of Illinois at Chicago\\Chicago, 
Illinois 60607\\licht@uic.edu}

\begin{abstract}
We consider the unparticle action that is made gauge invariant by 
inclusion of an open Wilson line factor.  In deriving vertexes from 
such an action it has been customary to use a form of differentiating the 
Wilson line originally proposed by Mandelstam. 
Using a simple example, we show that the Mandelstam derivative is 
mathematically inconsistent. We show that there are two ways 
to define differentiation of the Wilson line. The mathematically 
consistent method is to differentiate the explicit dependence of the 
line on the endpoint.  The other method is a functional derivative 
and corresponds in a limiting case to the Mandelstam derivative.  We also show 
that the only path that can be used in the Wilson line integral that leaves the 
unparticle action both Poincare and scale invariant is the straight 
line.
\end{abstract}

\maketitle

\section{1. Introduction}\label{S:intro}

The scalar bosonic unparticle action introduced by Georgi~\cite{Geo-1}~\cite{Geo-2} has been 
modified by Terning et al~\cite{Tern-1}~\cite{Tern-2} to include gauge field interactions. The 
resulting bosonic unparticle action is, apart from constants,

\begin{equation}\label{E:BUA}
I = \int {d^4 xd^4 y} \Phi _u^\dag  \left( x \right)K\left( {x,y} \right)W_\Lambda  \left( {x,y} \right)\Phi _u \left( y \right)
\end{equation}
where the K is determined as the inverse of the propagator,

\begin{equation}
K\left( {x,y} \right) = \left[ { - \left( {\partial _\mu  \partial ^\mu  } \right)_x  - i\varepsilon } \right]^{2 - d_u } \delta ^4 \left( {x - y} \right)
\end{equation}

and the $W_{\Lambda}$ is a path ordered Wilson line, introduced to make the action gauge invariant:

\begin{equation}
W_\Lambda  \left( {x,y} \right) = P\exp \left[ { - ig\int_y^x {A_\alpha  \left( \zeta  \right)d\zeta ^\alpha  } } \right]
\end{equation}
The symbol $\Lambda$ indicates the particular path chosen between x and y.  
The symbol P indicates the path ordering, from y on the right to x on the left.  
Choosing a path amounts to finding a vector valued function of the position 
vectors x, y and of a parameter $\lambda$, $\zeta ^\alpha  \left( {x,y,\lambda } \right)$
 , such that

\begin{equation}\label{E:ZLIM}
\begin{gathered}
  \zeta ^\alpha  \left( {x,y,0} \right) = y^\alpha   \\ 
  \zeta ^\alpha  \left( {x,y,1} \right) = x^\alpha   \\ 
\end{gathered} 
\end{equation}

The integral in Eq. (~\ref{E:BUA}) cannot be evaluated without an explicit choice of the 
function $\zeta$  . In calculating Eq. (~\ref{E:BUA}) it can be necessary to find the derivatives of 
the Wilson line.  .   In much of the current literature~\cite{Tern-1},~\cite{Tern-2},~\cite{Liao-1},~\cite{Liao-2}, 
the Mandelstam condition~\cite{Mand} is used, which defines the derivative as 

\begin{equation}\label{E:MD}
\frac{\partial }
{{\partial x^\nu  }}W_\zeta  \left( {x,y} \right) =  - igA_\nu  \left( x \right)W_\zeta  \left( {x,y} \right)
\end{equation}

There has been some controversy recently as to the validity of the Mandelstam 
condition~\cite{gms},~\cite{all-1},~\cite{all-3}.  In the following we will show in Section 2, 
using a simple example, that the use of the Mandelstam derivative in integrals 
similar to that in the unparticle action is indeed mathematically inconsistent.

In Section 3 we will give the appropriate form of Wilson line derivative to be 
used in the unparticle action.  It takes into account the effect on 
the path $\Lambda$ of the displacement of x to $\delta  x$ .  In Section 4 we 
show that the Mandelstam derivative is a special case of a functional derivative 
of the path function $\zeta$. It could conceivably be of use in some 
context, but not in evaluating the unparticle action.  

In Section 5 we show that the requirement that the unparticle action be both 
Lorentz and scale invariant requires that the path be the straight line between 
the points x and y.

\section{2. The Mandelstam Derivative}

We consider in this section a simplified form of Eq. (~\ref{E:BUA}) that can be evaluated 
either by ordinary integration, or by partial integration.  The partial 
integration involves differentiating a Wilson line which we may do using either 
ordinary derivatives or Mandelstam derivatives.  We will show that using the 
ordinary derivative in the partial integration gives the same result as does 
ordinary integration, but using the Mandelstam derivative does not.

We take as the gauge field A the electromagnetic vector potential in the presence 
of a uniform magnetic field B: 

\begin{equation}
\vec A =  - \frac{1}
{2}\vec r \times \vec B
\end{equation}

The field is abelian, so no path ordering is required. For the path we will take 
the straight line between x and y, parametrised as

\begin{equation}
\vec r\left( \lambda  \right) = \lambda \vec x + \left( {1 - \lambda } \right)\vec y
\end{equation}

Then

\begin{equation}
d\vec r = \left( {\vec x - \vec y} \right)d\lambda 
\end{equation}

and

\begin{equation}
\begin{gathered}
  \int_y^x {} \vec A \cdot d\vec r = \int_0^1 {} \frac{{ - \left( {\lambda \vec x + \left( {1 - \lambda } \right)\vec y} \right)}}
{2} \times \vec B \cdot \left( {\vec x - \vec y} \right)d\lambda  \\ 
   =  - \frac{1}
{2}\vec x \times \vec y \cdot \vec B \\ 
\end{gathered} 
\end{equation}

Then the Wilson line is

\begin{equation}
W\left( {x,y} \right) = \exp \left[ {i\frac{g}
{2}\vec x \times \vec y \cdot \vec B} \right]
\end{equation}

One way to take the derivative of this is to differentiate the explicit x 
dependence, then

\begin{equation}
\begin{gathered}
  \vec \nabla _x W\left( {x,y} \right) = \vec \nabla _x \exp \left[ {i\frac{g}
{2}\vec x \times \vec y \cdot \vec B} \right] \\ 
   = i\frac{g}
{2}\vec y \times \vec B\exp \left[ {i\frac{g}
{2}\vec x \times \vec y \cdot \vec B} \right] \\ 
\end{gathered} 
\end{equation}

It is also possible to use the Mandelstam derivative $\vec D_x^M $
  , defined as

\begin{equation}
\vec D_x^M W\left( {\vec x,\vec y} \right) =  - ig\vec A\left( {\vec x} \right)\exp \left[ { - ig\int_y^x {\vec A\left( {\vec r} \right) \cdot d\vec r} } \right]
\end{equation}

Then

\begin{equation}
\vec D_x^M W\left( {x,y} \right) = i\frac{g}
{2}\vec x \times \vec B\exp \left[ {i\frac{g}
{2}\vec x \times \vec y \cdot \vec B} \right]
\end{equation}

We take B to be normal to a plane, and consider the integral over the plane,

\begin{equation}
I\left( {a,b} \right) = \int {d^2 xd^2 y} \Psi _a \left( x \right)W\left( {x,y} \right)\Phi _b \left( y \right)
\end{equation}

Where

\begin{equation}
\begin{gathered}
  \Psi _a \left( x \right) = \nabla _x^2 \exp \left[ { - a\frac{{x^2 }}
{2}} \right] \\ 
  \Phi _b \left( y \right) = \exp \left[ { - b\frac{{y^2 }}
{2}} \right] \\ 
\end{gathered} 
\end{equation}

We evaluate this integral three ways:

(1)  A simple straightforward integration, without integrating by parts.  We 
call the result $I\left( {a,b} \right)$
 .

(2)  Integrating by parts, so that the derivatives act on $W\left( {a,b} \right)$
 , but differentiating the explicit dependence of  $W\left( {a,b} \right)$
 on x,  using the operator $\nabla _x^2 $.  We call this integral $I_\nabla  \left( {a,b} \right)$.

(3)  Integrating by parts, but using the Mandelstam operator $\vec D_x^M $
  when the derivative acts on $W\left( {a,b} \right)$ .  We call the result $I_D \left( {a,b} \right)$
  .  

The result is

\begin{equation}
I\left( {a,b} \right) = I_\nabla  \left( {a,b} \right) =  - 2\frac{{\left( {\pi gB} \right)^2 }}
{{a\left( {b + \frac{{\left( {gB} \right)^2 }}
{{4a}}} \right)^2 }}
\end{equation}

but

\begin{equation}
I_D \left( {a,b} \right) =  - 2\frac{{\left( {\pi gB} \right)^2 b}}
{{a^2 \left( {b + \frac{{\left( {gB} \right)^2 }}
{{4a}}} \right)^2 }}
\end{equation}

In the general case, when $a \ne b$
 ,  we have

\begin{equation}
I_D \left( {a,b} \right) \ne I\left( {a,b} \right)
\end{equation}

It should be obvious that  $I\left( {a,b} \right)$ is the correct value of the 
integral.  Integrating by parts should not change the value of an integral.  We 
have to conclude that the Mandelstam derivative is not the appropriate way to 
differentiate the Wilson line in a non-local action integral.

\section{3.  Explicit differentiation}

One way of differentiating the Wilson line is to differentiate with respect to 
one of the two endpoints x or y, keeping the other endpoint fixed, and also 
keeping the function $\zeta$ fixed.  Writing

\begin{equation}
W_\Lambda  \left( {x,y} \right) = P\exp \left[ { - ig\int_0^1 {A_\alpha  \left( \zeta  \right)\frac{{d\zeta ^\alpha  }}
{{d\lambda }}d\lambda } } \right]
\end{equation}

Defining

\begin{equation}
W_\Lambda  \left( {x,\zeta \left( \lambda  \right)} \right) = P\exp \left[ { - ig\int_\lambda ^1 {A_\alpha  \left( \zeta  \right)\frac{{d\zeta ^\alpha  }}
{{d\lambda '}}d\lambda '} } \right]
\end{equation}

\begin{equation}
W_\Lambda  \left( {\zeta \left( \lambda  \right),y} \right) = P\exp \left[ { - ig\int_0^\lambda  {A_\alpha  \left( \zeta  \right)\frac{{d\zeta ^\alpha  }}
{{d\lambda '}}d\lambda '} } \right]
\end{equation}

We have

\begin{equation}\label{E:DW1}
\frac{{\partial W_\Lambda  \left( {x,y} \right)}}
{{\partial x^\mu  }} =  - ig\int_0^1 {W_\Lambda  \left( {x,\zeta \left( \lambda  \right)} \right)} \frac{\partial }
{{\partial x^\mu  }}\left( {A_\alpha  \left( \zeta  \right)\frac{{d\zeta ^\alpha  }}
{{d\lambda }}} \right)W_\Lambda  \left( {\zeta \left( \lambda  \right),y} \right)d\lambda 
\end{equation}

Using

\begin{equation}\label{E:DW2}
\frac{\partial }
{{\partial x^\mu  }}\left( {A_\alpha  \left( \zeta  \right)\frac{{d\zeta ^\alpha  }}
{{d\lambda }}} \right) = \frac{{\partial A_\alpha  \left( \zeta  \right)}}
{{\partial \zeta ^\beta  }}\frac{{\partial \zeta ^\beta  }}
{{\partial x^\mu  }}\frac{{d\zeta ^\alpha  }}
{{d\lambda }} + A_\alpha  \left( \zeta  \right)\frac{d}
{{d\lambda }}\frac{{\partial \zeta ^\alpha  }}
{{\partial x^\mu  }}
\end{equation}

we get, integrating by parts,

\begin{equation}\label{E:DW3}
\begin{gathered}
  \int_0^1 {W_\Lambda  \left( {x,\zeta \left( \lambda  \right)} \right)} A_\alpha  \left( \zeta  \right)\frac{d}
{{d\lambda }}\frac{{\partial \zeta ^\alpha  }}
{{\partial x^\mu  }}W_\Lambda  \left( {\zeta \left( \lambda  \right),y} \right)d\lambda  = W_\Lambda  \left( {x,\zeta \left( \lambda  \right)} \right)A_\alpha  \left( \zeta  \right)\frac{{\partial \zeta ^\alpha  }}
{{\partial x^\mu  }}\left. {W_\Lambda  \left( {\zeta \left( \lambda  \right),y} \right)} \right|_0^1  \hfill \\
  \quad \quad \quad \quad \quad \quad  - \int_0^1 {W_\Lambda  \left( {x,\zeta \left( \lambda  \right)} \right)} \left\{ {\frac{{\partial A_\alpha  \left( \zeta  \right)}}
{{\partial \zeta ^\beta  }} + ig\left[ {A_\beta  \left( \zeta  \right),A_\alpha  \left( \zeta  \right)} \right]} \right\}\frac{{d\zeta ^\beta  }}
{{d\lambda }}\frac{{\partial \zeta ^\alpha  }}
{{\partial x^\mu  }}W_\Lambda  \left( {\zeta \left( \lambda  \right),y} \right)d\lambda  \hfill \\ 
\end{gathered} 
\end{equation}

In differentiating by x we are holding y fixed, so Eqs. (~\ref{E:ZLIM}) imply that

\begin{equation}
\begin{gathered}
  \frac{{\partial \zeta ^\gamma  \left( {x,y,1} \right)}}
{{\partial x^\mu  }} = \frac{{\partial x^\gamma  }}
{{\partial x^\mu  }} = \delta _\mu ^\gamma   \hfill \\
  \frac{{\partial \zeta ^\gamma  \left( {x,y,0} \right)}}
{{\partial x^\mu  }} = \frac{{\partial y^\gamma  }}
{{\partial x^\mu  }} = 0 \hfill \\ 
\end{gathered} 
\end{equation}

Also using

\begin{equation}
\begin{gathered}
  W_\Lambda  \left( {x,\zeta \left( 1 \right)} \right) = W_\Lambda  \left( {x,x} \right) = 1 \hfill \\
  W_\Lambda  \left( {\zeta \left( 1 \right),y} \right) = W_\Lambda  \left( {x,y} \right) \hfill \\ 
\end{gathered} 
\end{equation}

and combining Eqs. (~\ref{E:DW1}), (~\ref{E:DW2}) and (~\ref{E:DW3}), we get

\begin{equation}\label{E:DWEXP}
\frac{\partial }
{{\partial x^\mu  }}W_\Lambda  \left( {x,y} \right) =  - igA_\mu  \left( x \right)W_\Lambda  \left( {x,y} \right) + ig\int_0^1 {} W_\Lambda  \left( {x,\zeta \left( \lambda  \right)} \right)F_{\beta \alpha } \left( {\zeta \left( \lambda  \right)} \right)\frac{{\partial \zeta ^\alpha  }}
{{\partial x^\mu  }}\frac{{d\zeta ^\beta  }}
{{d\lambda }}W_\Lambda  \left( {\zeta \left( \lambda  \right),y} \right)d\lambda 
\end{equation}

where $F_{\beta \alpha}$ is the field strength:

\begin{equation}
F_{\beta \alpha }  = \partial _\beta  A_\alpha   - \partial _\alpha  A_\beta   + ig\left[ {A_\beta  ,A_\alpha  } \right]
\end{equation}

For a continuous non-null path, the second term on the right hand side of 
Eq. (~\ref{E:DWEXP}) can only be eliminated if the field strength is zero.

\section{4. The Functional Derivative}

 Mandelstam however takes for the derivative only the first term on the right 
 hand side of Eq. (~\ref{E:DWEXP}):
 
\begin{equation}
\frac{\partial }
{{\partial x^\nu  }}W_\Lambda  \left( {x,y} \right) =  - igA_\nu  \left( x \right)W_\Lambda  \left( {x,y} \right)
\end{equation}

The argument for this result is essentially the idea that the derivative can be 
taken by defining it as

\begin{equation}
\frac{\partial }
{{\partial x^\nu  }}W_\Lambda  \left( {x,y} \right) = \lim _{\Delta x \to 0} \frac{{W_{\Lambda '} \left( {x + \Delta x,y} \right) - W_\Lambda  \left( {x,y} \right)}}
{{\Delta x^\nu  }}
\end{equation}

where $\Lambda '$ is a path that is identical with $\Lambda$ up to the point x, 
where it goes off to x + $\Delta$x.  This is a change in the form of the path 
function $\zeta$.  We will show that it can be expressed in terms of a 
functional derivative. 

Consider the vector tangent to the path defined as

\begin{equation}
\eta ^\alpha  \left( \lambda  \right) = \frac{{d\zeta ^\alpha  \left( \lambda  \right)}}
{{d\lambda }}
\end{equation}

We can define a functional derivative with respect to this vector:

\begin{equation}
\frac{{\delta \eta ^\alpha  \left( \lambda  \right)}}
{{\delta \eta ^\beta  \left( {\lambda '} \right)}} = \delta _\beta ^\alpha  \delta \left( {\lambda  - \lambda '} \right)
\end{equation}

now the path function can be written as

\begin{equation}
\begin{gathered}
  \zeta ^\alpha  \left( \lambda  \right) = y^\alpha   + \int_0^\lambda  {\eta ^\alpha  \left( {\lambda '} \right)} d\lambda ' \\ 
   = x^\alpha   - \int_\lambda ^1 {\eta ^\alpha  \left( {\lambda '} \right)} d\lambda ' \\ 
\end{gathered} 
\end{equation}

which leads to

\begin{equation}
\begin{gathered}
  \left. {\frac{{\delta \zeta ^\alpha  \left( \lambda  \right)}}
{{\delta \eta ^\beta  \left( {\lambda _0 } \right)}}} \right|_y  =  + \delta _\beta ^\alpha  \theta \left( {\lambda  - \lambda _0 } \right) \hfill \\
  \left. {\frac{{\delta \zeta ^\alpha  \left( \lambda  \right)}}
{{\delta \eta ^\beta  \left( {\lambda _0 } \right)}}} \right|_x  =  - \delta _\beta ^\alpha  \theta \left( {\lambda _0  - \lambda } \right) \hfill \\ 
\end{gathered} 
\end{equation}

The functional derivative of the Wilson line, holding y fixed, is

\begin{equation}\label{E:FDW1}
\left. {\frac{{\delta W_\Lambda  \left( {x,y} \right)}}
{{\delta \eta ^\beta  \left( \lambda  \right)}}} \right|_y  =  - ig\int_0^1 {W_\Lambda  \left( {x,\zeta \left( {\lambda '} \right)} \right)\left. {\frac{\delta }
{{\delta \eta ^\beta  \left( \lambda  \right)}}\left( {A_\alpha  \left( {\zeta \left( {\lambda '} \right)} \right)\eta ^\alpha  \left( {\lambda '} \right)} \right)} \right|_y W_\Lambda  \left( {\zeta \left( {\lambda '} \right),y} \right)} d\lambda '
\end{equation}

Expanding,

\begin{equation}
\begin{gathered}
  \left. {\frac{\delta }
{{\delta \eta ^\beta  \left( \lambda  \right)}}\left( {A_\alpha  \left( {\zeta \left( {\lambda '} \right)} \right)\eta ^\alpha  \left( {\lambda '} \right)} \right)} \right|_y  = \frac{{\partial A_\alpha  \left( {\zeta \left( {\lambda '} \right)} \right)}}
{{\partial \zeta ^\gamma  \left( {\lambda '} \right)}}\left. {\frac{{\delta \zeta ^\gamma  \left( {\lambda '} \right)}}
{{\delta \eta ^\beta  \left( \lambda  \right)}}} \right|_y \eta ^\alpha  \left( {\lambda '} \right) + A_\alpha  \left( {\zeta \left( {\lambda '} \right)} \right)\frac{{\delta \eta ^\alpha  \left( {\lambda '} \right)}}
{{\delta \eta ^\beta  \left( \lambda  \right)}} \\ 
   = \frac{{\partial A_\alpha  \left( {\zeta \left( {\lambda '} \right)} \right)}}
{{\partial \zeta ^\gamma  \left( {\lambda '} \right)}}\delta _\beta ^\gamma  \theta \left( {\lambda ' - \lambda } \right)\eta ^\alpha  \left( {\lambda '} \right) + A_\alpha  \left( {\zeta \left( {\lambda '} \right)} \right)\delta _\beta ^\alpha  \delta \left( {\lambda  - \lambda '} \right) \\ 
\end{gathered} 
\end{equation}

Then Eq. (~\ref{E:FDW1}) becomes

\begin{equation}\label{E:FDW2}
\begin{gathered}
  \left. {\frac{{\delta W_\Lambda  \left( {x,y} \right)}}
{{\delta \eta ^\beta  \left( \lambda  \right)}}} \right|_y  =  - ig\int_\lambda ^1 {W_\Lambda  \left( {x,\zeta \left( {\lambda '} \right)} \right)\frac{{\partial A_\alpha  \left( {\zeta \left( {\lambda '} \right)} \right)}}
{{\partial \zeta ^\beta  \left( {\lambda '} \right)}}\eta ^\alpha  \left( {\lambda '} \right)W_\Lambda  \left( {\zeta \left( {\lambda '} \right),y} \right)} d\lambda ' \hfill \\
  \quad \quad \quad \quad \quad \quad \quad  - igW_\Lambda  \left( {x,\zeta \left( \lambda  \right)} \right)A_\beta  \left( {\zeta \left( {\lambda '} \right)} \right)W_\Lambda  \left( {\zeta \left( \lambda  \right),y} \right) \hfill \\ 
\end{gathered} 
\end{equation}

If we go to the limit, $\lambda  \to 1$ , the first term on the right hand side 
of Eq. (~\ref{E:FDW2}) vanishes, and we are left with

\begin{equation}
\left. {\frac{{\delta W_\Lambda  \left( {x,y} \right)}}
{{\delta \eta ^\beta  \left( 1 \right)}}} \right|_y  =  - igA_\beta  \left( x \right)W_\Lambda  \left( {x,y} \right)
\end{equation}

and similiarly,

\begin{equation}
\left. {\frac{{\delta W_\Lambda  \left( {x,y} \right)}}
{{\delta \eta ^\beta  \left( 0 \right)}}} \right|_x  =  + igW_\Lambda  \left( {x,y} \right)A_\beta  \left( y \right)
\end{equation}

These are very similar to the Mandelstam conditions.  It is tempting therefore 
to define a kind of end point derivative as

\begin{equation}
\begin{gathered}
  \frac{{DW_\Lambda  \left( {x,y} \right)}}
{{Dx^\mu  }} = \left. {\frac{{\delta W_\Lambda  \left( {x,y} \right)}}
{{\delta \eta ^\mu  \left( 1 \right)}}} \right|_y  \hfill \\
  \frac{{DW_\Lambda  \left( {x,y} \right)}}
{{Dy^\mu  }} = \left. {\frac{{\delta W_\Lambda  \left( {x,y} \right)}}
{{\delta \eta ^\mu  \left( 0 \right)}}} \right|_x  \hfill \\ 
\end{gathered} 
\end{equation}

The problem arises in identifying this  $D/Dx^\mu  $ with the  
$\partial /\partial x^\mu  $ that appears in Eq. (~\ref{E:MD}).  The 
operation $\partial /\partial x^\mu  $ is simple differentiation of the explicit 
dependence on x. It is hard to see how it could involve a change in the function 
that defines that explicit dependence.  The operator $D/Dx^\mu  $ is the limit 
of a functional derivative. It is defined only on functionals of the path 
defining function $\zeta ^\alpha  \left( \lambda  \right)$. It could not be 
applied to the unparticle field $\Phi _u \left( x \right)$.

\section{5. Scale Invariance and the Straight Line}

With no gauge fields, the Bosonic unparticle action is, apart from an overall 
constant:

\begin{equation}\label{E:UNGA}
I_u  = \int {d^4 xd^4 y} \Phi _u^\dag  \left( x \right)K\left( {x - y} \right)\Phi _u \left( y \right)
\end{equation}

where

\begin{equation}
K\left( {x - y} \right) = \int {\frac{{d^4 k}}
{{\left( {2\pi } \right)^4 }}} \left( {k^2  - i\varepsilon } \right)^{2 - d_u } e^{ - ik \cdot \left( {x - y} \right)} 
\end{equation}

K is a Poincare invariant function, and with 

\begin{equation}
U\left( {\Lambda ,b} \right)\Phi _u \left( x \right)U^\dag  \left( {\Lambda ,b} \right) = \Phi _u \left( {\Lambda x + b} \right)
\end{equation}

the action is invariant under the Poincare group.  

$I_{u}$ is also scale invariant.  With the scale transformation defined as

\begin{equation}
U\left( a \right)\Phi _u \left( x \right)U^\dag  \left( a \right) = a^{d_u } \Phi _u \left( {ax} \right)
\end{equation}

We get

\begin{equation}
U\left( a \right)I_u U^\dag  \left( a \right) = a^{2d_u } \int {d^4 xd^4 y} \Phi _u^\dag  \left( {ax} \right)K\left( {x - y} \right)\Phi _u \left( {ay} \right)
\end{equation}

Letting

\begin{equation}\label{E:CCH}
\begin{gathered}
  x \to x'/a \hfill \\
  y \to y'/a \hfill \\ 
\end{gathered}
\end{equation}
and using

\begin{equation}
K\left( {\frac{z}
{a}} \right) = a^{8 - 2d_u } K\left( z \right)
\end{equation}

we get

\begin{equation}
U\left( a \right)I_u U^\dag  \left( a \right) = \int {d^4 x'd^4 y'} \Phi _u^\dag  \left( {x'} \right)K\left( {x' - y'} \right)\Phi _u \left( {y'} \right) = I_u 
\end{equation}

To make the action gauge invariant we can include a Wilson line as in 
Eq. (~\ref{E:BUA}), involving an integral over a path $\Lambda$. 
We will show here that the requirement that $I_{u}$  be both Poincare and 
scale invariant implies that the path $\Lambda$ can only be the straight line 
connecting x and y.

The only two vectors in the integral are $x^{\mu}$ and $y^{\mu}$. Lorentz 
invariance then implies that $\zeta ^\mu  \left( {x,y,\lambda } \right)$ 
must be of the form

\begin{equation}\label{E:ZFG}
\zeta ^\mu  \left( {x,y,\lambda } \right) = f\left( {u,v,w,\lambda } \right)x^\mu   + g\left( {u,v,w,\lambda } \right)y^\mu  
\end{equation}

where
\begin{equation}
\begin{gathered}
  u = x \cdot x \\ 
  v = x \cdot y \\ 
  w = y \cdot y \\ 
\end{gathered} 
\end{equation}

and

\begin{equation}
\begin{gathered}
  f\left( {u,v,w,0} \right) = 0\quad ,\quad f\left( {u,v,w,1} \right) = 1 \hfill \\
  g\left( {u,v,w,0} \right) = 1\quad ,\quad g\left( {u,v,w,1} \right) = 0 \hfill \\ 
\end{gathered} 
\end{equation}

We take  $A_\mu  \left( z \right)$ to be a zero mass field, which can be shown 
to scale as

\begin{equation}
U\left( a \right)A_\mu  \left( z \right)U^\dag  \left( a \right) = aA_\mu  \left( {az} \right)
\end{equation}

with dimension 1.   The line integral now scales as

\begin{equation}
U\left( a \right)\int_0^1 {d\lambda {\kern 1pt} } A_\mu  \left( {\zeta \left( {x,y,\lambda } \right)} \right)\frac{{d\zeta ^\mu  \left( {x,y,\lambda } \right)}}
{{d\lambda }}U^\dag  \left( a \right) = a\int_0^1 {d\lambda {\kern 1pt} } A_\mu  \left( {a\zeta \left( {x,y,\lambda } \right)} \right)\frac{{d\zeta ^\mu  \left( {x,y,\lambda } \right)}}
{{d\lambda }}
\end{equation}

Making the coordinate change of Eq. (~\ref{E:CCH}), we see that scale invariance of 
the line integral requires that

\begin{equation}
a\zeta ^\mu  \left( {\frac{{x'}}
{a},\frac{{y'}}
{a},\lambda } \right) = \zeta ^\mu  \left( {x',y',\lambda } \right)
\end{equation}

The form given for $\zeta ^\mu  \left( {x,y,\lambda } \right)$  in 
Eq. (~\ref{E:ZFG}) then implies that

\begin{equation}
\begin{gathered}
  f\left( {a^{ - 2} u',a^{ - 2} v',a^{ - 2} w',\lambda } \right) = f\left( {u',v',w',\lambda } \right) \\ 
  g\left( {a^{ - 2} u',a^{ - 2} v',a^{ - 2} w',\lambda } \right) = g\left( {u',v',w',\lambda } \right) \\ 
\end{gathered} 
\end{equation}

Differentiating with respect to $a^{ - 2} $  at $a = 1$ then gives

\begin{equation}\label{E:UDUFG}
\begin{gathered}
  \left( {u\frac{\partial }
{{\partial u}} + v\frac{\partial }
{{\partial v}} + w\frac{\partial }
{{\partial w}}} \right)f\left( {u,v,w,\lambda } \right) = 0 \\ 
  \left( {u\frac{\partial }
{{\partial u}} + v\frac{\partial }
{{\partial v}} + w\frac{\partial }
{{\partial w}}} \right)g\left( {u,v,w,\lambda } \right) = 0 \\ 
\end{gathered} 
\end{equation}

We now look at the consequences of translational invariance.  Under translation 
through a vector $b^\mu  $ , the line integral becomes

\begin{equation}
U\left( {1,b} \right)\int_0^1 {} A_\mu  \left( {\zeta \left( {x,y,\lambda } \right)} \right)\frac{{d\zeta ^\mu  \left( {x,y,\lambda } \right)}}
{{d\lambda }}U^\dag  \left( {1,b} \right)d\lambda  = \int_0^1 {} A_\mu  \left( {\zeta \left( {x,y,\lambda } \right) + b} \right)\frac{{d\zeta ^\mu  \left( {x,y,\lambda } \right)}}
{{d\lambda }}d\lambda 
\end{equation}

The ungauged action of Eq. (~\ref{E:UNGA}) under such a transformation is restored to its 
original form by the change of variables:

\begin{equation}
\begin{gathered}
  x^\mu   \to x^{\mu\prime} - b^\mu   \hfill \\
  y^\mu   \to y^{\mu\prime} - b^\mu   \hfill \\ 
\end{gathered} 
\end{equation}

The line integral then goes into its original form if and only if

\begin{equation}\label{E:FGb}
\begin{gathered}
  f\left( {u' - 2x' \cdot b + b^2 ,v' - x' \cdot b - y' \cdot b + b^2 
  ,w' - 2y' \cdot b + b^2 ,\lambda } \right)\left( {x^{\mu\prime} - b^\mu  } \right) +  \hfill \\
  g\left( {u' - 2x' \cdot b + b^2 ,v' - x' \cdot b - y' \cdot b + b^2 
  ,w' - 2y' \cdot b + b^2 ,\lambda } \right)\left( {y^{\mu\prime} - b^\mu  } \right) \hfill \\
  \quad \quad \quad \quad  = f\left( {u',v',w',\lambda } 
  \right)x^{\mu\prime} + g\left( {u',v',w',\lambda } 
  \right)y^{\mu\prime} - b^\mu   \hfill \\ 
\end{gathered} 
\end{equation}

Differentiating Eq. (~\ref{E:FGb}) with respect to $b^\alpha  $  at $b = 0$  yields, 
after dropping the primes,

\begin{equation}\label{E:DFB}
\begin{gathered}
  x^\mu  \left( {2x_\alpha  \partial _u  + 2y_\alpha  \partial _w  + \left( {x_\alpha   + y_\alpha  } \right)\partial _v } \right)f + y^\mu  \left( {2x_\alpha  \partial _u  + 2y_\alpha  \partial _w  + \left( {x_\alpha   + y_\alpha  } \right)\partial _v } \right)g \hfill \\
  \quad \quad \quad  + \left( {f + g} \right)\delta _\alpha ^\beta   - \delta _\alpha ^\beta   = 0 \hfill \\ 
\end{gathered} 
\end{equation}

The tensors $x^\mu  x_\alpha  \,,\,y^\mu  y_\alpha  \,,\,x^\mu  y_\alpha  \,,\,y^\mu  x_\alpha  \,,\,\delta _\alpha ^\mu  $
are linearly independent.  In Eq. (~\ref{E:DFB}) their coefficients must each add up 
to zero.  This yields
  
\begin{equation}\label{E:DUWFG}
\begin{gathered}
  \left( {2\partial _u  + \partial _v } \right)f = 0 \\ 
  \left( {2\partial _w  + \partial _v } \right)f = 0 \\ 
  \left( {2\partial _u  + \partial _v } \right)g = 0 \\ 
  \left( {2\partial _w  + \partial _v } \right)g = 0 \\ 
\end{gathered} 
\end{equation}

and

\begin{equation}\label{E:SFG}
f + g = 1
\end{equation}

Eqs. (~\ref{E:DUWFG}, a and b) imply that

\begin{equation}\label{E:DUDWF}
\partial _u f = \partial _w f
\end{equation}

Then using this and Eq. (~\ref{E:DUWFG},a ) in Eq. (~\ref{E:UDUFG}, a) we have

\begin{equation}
\left( {u - 2v + w} \right)\partial _u f = \left( {x - y} \right)^2 \partial _u f = 0
\end{equation}

With $x \ne y$  we get f independent of u, then Eqs. 
(~\ref{E:DUWFG}) and (~\ref{E:DUDWF}) imply 
that f is also independent of v and w, therefore it depends only on 
$\lambda$ .  Similarly, we find that g also depends only on $\lambda$ .  
Defining a new line parameter,

\begin{equation}
\lambda ' = f\left( \lambda  \right)
\end{equation}

we see, using Eq. (~\ref{E:SFG}) and Eq. (~\ref{E:ZFG}), that

\begin{equation}
\zeta ^\mu  \left( {x,y,\lambda } \right) = \lambda 'x^\mu   + \left( {1 - \lambda '} \right)y^\mu  
\end{equation}

the equation for a straight line connecting x and y.

\section{6. Conclusion}

We have shown that the Mandelstam derivative of the open Wilson line leads to 
mathematically inconsistent results in the unparticle action.  It is not the 
same as the ordinary derivative, but it is a special case of a functional 
derivative of the Wilson line.  We give explicit expressions for the ordinary 
derivative of the Wilson line and also for the general functional derivative. 
Lastly, we have shown that the combination of Poincare and scale invariance for 
the gauged unparticle action require that the path in the Wilson line integral 
be a straight line.

The interaction of an unparticle and a gauge field with a straight line Wilson 
integral has been investigated in Ref.~\cite{all-1}.  It leads to rather complicated 
expressions for the unparticle-gauge field vertexes. In 
Ref.~\cite{all-2} however it was shown that it is possible to construct a gauge invariant unparticle action 
without the use of the Wilson line integral.

\section{Acknowledgements}\label{S:Acknow}
I would like to thank Wai-Yee Keung for interesting me in this 
subject.  I would also like to thank Galloway, Martin and Stancato 
for their very interesting comments.

\end{document}